\documentclass[%
article,
 floatfix,
twocolumn,
 aps,
 prb,
 groupedaddress,
 amsmath,amssymb,
floatfix,
]{revtex4-1}

\usepackage{float}
\usepackage{graphicx}% Include Fig. files
\usepackage{dcolumn}% Align table columns on decimal point
\usepackage{bm}% bold math
\usepackage{siunitx}
\usepackage{epstopdf}
\usepackage{color}
\usepackage{lipsum}
\usepackage{bm}% bold math
\usepackage{titlesec}

\usepackage{polski}
\usepackage{epstopdf}
\usepackage{multirow}
\usepackage{siunitx}
\usepackage{color}
\usepackage{verbatim}
\usepackage{lipsum}
\usepackage[english]{babel}
\usepackage[T1]{fontenc}
\usepackage[utf8]{inputenc}
\usepackage{hyperref}% add hypertext capabilities
\usepackage[mathlines]{lineno}% Enable numbering of text and display math
\addto\captionsenglish{}
\begin{document}

\preprint{PRB 06.05.2019}

\title{Formation of N\'eel Type Skyrmions in an Antidot Lattice with Perpendicular Magnetic Anisotropy }% Force line breaks with \\

\author{S.~Saha$^{1,2}$}
\email[]{Experimental part: susmita.saha@psi.ch}
\author{M.~Zelent$^{3}$ }
\email[]{Simulations part: mateusz.zelent@gmail.com}
\author{S.~Finizio$^{2}$ } 
\author{M.~Mruczkiewicz$^{4}$ } 
\author{S.~Tacchi$^{5}$ } 
\author{A. K. Suszka$^{1,2}$ } 
\author{S. Wintz$^{2,6}$ } 
\author{N. S. Bingham$^{1,2,7}$ }
\author{J. Raabe$^{2}$ } 
\author{M.~Krawczyk$^{4}$ } 
\author{L. J. Heyderman$^{1,2}$ } 

\affiliation{%
$^{1}$Laboratory for Mesoscopic Systems, Department of Materials, ETH Zurich, 8093 Zurich, Switzerland
}%
\affiliation{%
$^{2}$Paul Scherrer Institut, 5232 Villigen PSI, Switzerland
}%
\affiliation{%
$^{3}$Faculty of Physics, Adam Mickiewicz University in Poznan, Umultowska 85, Poznan, PL-61-614 Poland
}%
\affiliation{%
$^{4}$Institute of Electrical Engineering, Slovak Academy of Sciences, Dubravska Cesta 9, SK-841-04 Bratislava, Slovakia
}%
\affiliation{%
$^{5}$Istituto Officina dei Materiali del CNR (CNR-IOM),
Sede Secondaria di Perugia, c/o Dipartimento di Fisica e Geologia, Universit\`a di Perugia, I-06123 Perugia, Italy
}%
\affiliation{%
$^{6}$Helmholtz-Zentrum Dresden-Rossendorf, 01328 Dresden, Germany
}%
\affiliation{%
$^{7}$Department of Applied Physics, Yale University, New Haven CT 06511, USA
}%

\date{\today}% It is always \today, today,
 % but any date may be explicitly specified

\begin{abstract}
Magnetic skyrmions are particle-like chiral spin textures found in a magnetic film with out-of-plane anisotropy and are considered to be potential candidates as information carriers in next generation data storage devices. Despite intense research into the nature of skyrmions and their dynamic properties, there are several key challenges that still need to be addressed. In particular, the outstanding issues are the reproducible generation, stabilization and confinement of skyrmions at room temperature. Here, we present a method for the capture of nanometer sized magnetic skyrmions in an array of magnetic topological defects in the form of an antidot lattice. With micromagnetic simulations, we elucidate the skyrmion formation in the antidot lattice and show that the capture is dependent on the antidot lattice parameters. This behavior is confirmed with scanning transmission x-ray microscopy measurements. This demonstration that a magnetic antidot lattice can be implemented as a host to capture skyrmions provides a new platform for experimental investigations of skyrmions and skyrmion based devices.
\end{abstract}
\maketitle

%\tableofcontents

\section{\label{sec:intro}Introduction}
In materials exhibiting perpendicular magnetic anisotropy (PMA), non-collinear spin textures can be stabilized in weak magnetic fields or even at remanence. In particular, labyrinth domains can shrink to form isolated magnetic skyrmions~\cite{Robler_Spontaneus_skyrmion_2006,Fert_skyrmions_on_track_2013}, which are topologically non-trivial stable spin textures characterized by a non-zero winding number with topological charge~\cite{Zhang2015TopologicalCommunication}. The antisymmetric exchange interaction, known as the Dzyaloshinskii-Moriya interaction (DMI)~\cite{boulle2016room}, which can arise due to large spin orbit coupling in the presence of the broken inversion symmetry in the material, helps to stabilize the chiral magnetic skyrmions. For asymmetric heavy-metal/ferromagnet/heavy-metal heterostructures, the inversion symmetry is broken by using different layers above and below the ferromagnetic layer, giving rise to an interfacial DMI.

\indent The Hamiltonian describing the DMI between two atomic spins $\textbf{S}_i$ and $\textbf{S}_j$ can be expressed as: $H_\mathrm{DM} = \textbf{-D}_\mathrm{DM} \cdot (\textbf{S}_{i} \times \textbf{S}_{j})$, where $\textbf{D}_\mathrm{DM}$ is the Dzyaloshinskii-Moriya vector. The direction of the DM vector is defined by the type of DMI in the material. For interfacial DMI, the orientation of the DM vector leads to the stabilization of N\'eel-type magnetic skyrmions~\cite{Chen2013} whereas the lack of structural inversion symmetry in bulk materials~\cite{Nagaosa_Topological_properties_and_dynamics_2013,Turgut2018EngineeringMagnets} (e.g. B20 materials) leads to the stabilization of Bloch-type magnetic skyrmions~\cite{Fert_Magnetic_skyrmions_applications_2017}. 
In addition, skyrmions can also be stabilized via dipolar interactions in systems with small or no DMI~\cite{Hrabec2014,PSSR:PSSR201700259,Finizio2018ThickSystems}. The magnetic domain structure stabilized in magnetic layers and multilayer stacks depends on the properties of the magnetic film and magnetic history of the sample, with the final spin configuration of the material determined by the ratio of the uniaxial anisotropy $K_\mathrm{u}$ to the demagnetization energy, the thickness of magnetic layer $d$, the strength of the DMI, and the number of layers present in the superlattice. By changing the orientation of the external magnetic field, one can obtain different domain structures during magnetization reversal, such as stripe, labyrinth, non-chiral bubble domains and, in some cases, skyrmions~\cite{Ng1995StabilityModel, Han_Skurmion_lattice_2010}.

Additionally, the magnetic properties of a film can be influenced via patterning. For example, magnonic crystals \cite{krawczyk2014review,Malago2015Spin-waveSpacer}, which are spatially modulated magnetic thin films, form well-defined spin wave band structures~\cite{ch} due to their periodicity. One common example of magnonic crystals are ferromagnetic antidot lattices~\cite{Pal2014,Tahir2017,1609.05663} consisting of periodically arranged holes in a ferromagnetic thin film. Antidot lattices have been proposed as a medium for data storage devices, where the information bit would be trapped between two consecutive antidots~\cite{Cowburn1997,Wang2006}. Additionally, antidot lattices patterned in magnetic multilayers with high PMA have also been considered as a candidates for perpendicular percolated media~\cite{Suess2006}, which constitutes an alternative approach to conventional bit patterned media. Antidots can be also employed to modify the intrinsic anisotropy of the system~\cite{Toporov2000LorentzArrays}. With regard to skyrmions, it has been shown that the edge of a magnetic element as well as defects~\cite{Reichhardt2015} can repel the skyrmions. It has also been shown that pinning sites can accelerate skyrmion motion~\cite{Reichhardt2015}. Therefore, an antidot lattice can be considered to be an array of artificial pinning sites where skyrmions are repelled from the edges of the holes. Thus, skyrmions will be trapped at the saddle point within the lattice, allowing for control of current driven motion. It has also been shown, with micromagnetic simulations,~\cite{Marchenko2015} that an antidot lattice can be tailored to host skyrmion-like bubbles, which form without the presence of DMI.

Here, we demonstrate that we can capture magnetic skyrmions within an antidot lattice fabricated in a [Pt (3 nm)/Co (1.1 nm)/Ta (4 nm)]$_{12}$ multilayer (ML) stack exhibiting both PMA and interfacial DMI (see Fig. 1). Imaging the nanoscale skyrmions using scanning transmission x-ray microscopy (STXM), we observe that magnetic skyrmions are created from a labyrinth domain pattern on application of a magnetic field. With micromagnetic simulations, we show that the control of formation of skyrmions within the antidot lattice can be tuned by varying the lattice constant of the array. For a compact lattice, no skyrmions are formed whereas, for a sparser one, skyrmions are formed within the antidot lattice. In particular, in both experiment and micromagnetic simulations, we find that that the basic criterion for the formation of skyrmions is to have open ends in the labyrinth domain structure, which do not form below a specific separation between the antidots edges. 

\section{Sample fabrication}

\begin{figure}[!bp]
\centering
\includegraphics[scale=1,width=\columnwidth]{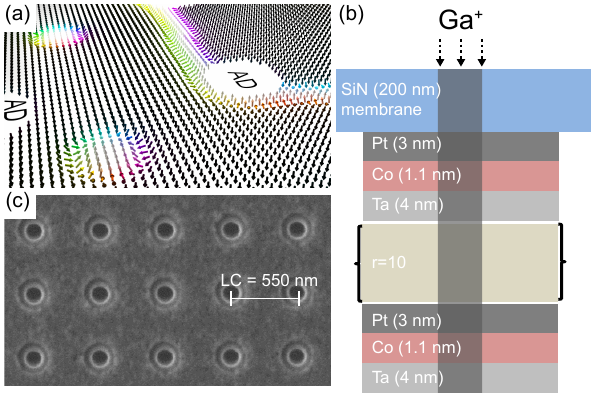} 
\caption{
(a) Visualization of the 3D spin orientation for skyrmions inside the antidot lattice in an applied magnetic field of
-35 mT. (b) Schematic of the multilayer stack with a fabricated antidot using focused Ga ion beam milling (FIB). (c) Scanning electron microscope image of the antidot lattice with antidot diameter $d$ = 250 nm and lattice constant LC = 550 nm patterned in the [Pt/Co/Ta]$_{12}$ film.
}
\label{fig1}
\end{figure}

Multilayer films of [Pt (3 nm)/ Co (1.1 nm)/ Ta (4 nm)]$_{12}$ were deposited on 200 nm thick x-ray transparent Si$_{3}$N$_{4}$ membranes supported by \SI{200}{\micro\metre} thick Si frames. The films were deposited using DC magnetron sputtering (AJA International Sputter System) with a confocal sputter up geometry at a base pressure of $2\times10^{-8}$ torr. The deposition of all of the layers was performed at a 3 mTorr Ar partial pressure. The antidot lattice was then patterned with focused ion beam milling (FIB) using a focused 30 kV $\mathrm{Ga}^{+}$ beam [Fig. 1(b)] with a beam current of 10 pA and a Gaussian spot size of 13 nm.
Not only does the incident $\mathrm{Ga}^{+}$ ion beam modify the magnetic properties of the film, but also the scattered $\mathrm{Ga}^{+}$ ions, resulting in a lateral distribution of the deposited dose. To reduce the effect of the scattered ions, the ion milling was carried out from the back of the membrane. In this case, the Si$_{3}$N$_{4}$ membrane absorbs most of the scattered $\mathrm{Ga}^{+}$ ions. Two different types of antidot lattices were fabricated with a hole (antidot) diameter $d$ of 250 nm and lattice constant (LC) of 550 and 1000 nm. A scanning electron micrograph of one of the patterned antidot lattices is shown in Fig. 1(c).

\section{Magnetic characterization}

The magnetic properties of the sputtered films were determined using various techniques. From the out-of-plane hysteresis loop, measured using a superconducting quantum interference device vibrating sample magnetometer (SQUID-VSM, Quantum Design MPMS3) and shown in Fig. 2(a), the saturation magnetization, ($M_\mathrm{s}$), was found to be $\approx 0.771*10^{6}$ A/m and saturation field $\approx 40$ mT. The interfacial DMI of the system was determined by measuring the non-reciprocal spin wave dispersion using Brillouin light scattering (BLS). The presence of the interfacial DMI induces a frequency difference, $\Delta f$, between oppositely propagating Damon-Eshbach modes. This exhibits a linear dependence on the spin wave vector $k$ and is proportional to the effective DMI constant $D$, according to the following relation:
\begin{equation}\label{eq:dmi_eq}
\Delta f = \frac{2 \gamma D}{\pi M_\mathrm{s}}k
\end{equation}
where $\gamma$ is the gyromagnetic ratio. The spin wave non-reciprocity, $\Delta f$, was determined by measuring the frequency difference between the Stokes (negative frequencies) and anti-Stokes (positive frequencies) peaks in the BLS spectra, which correspond to spin waves propagating in the opposite directions~\cite{Tacchi2017InterfacialThickness}.

\begin{figure}
\includegraphics[scale=1, width=.9\columnwidth]{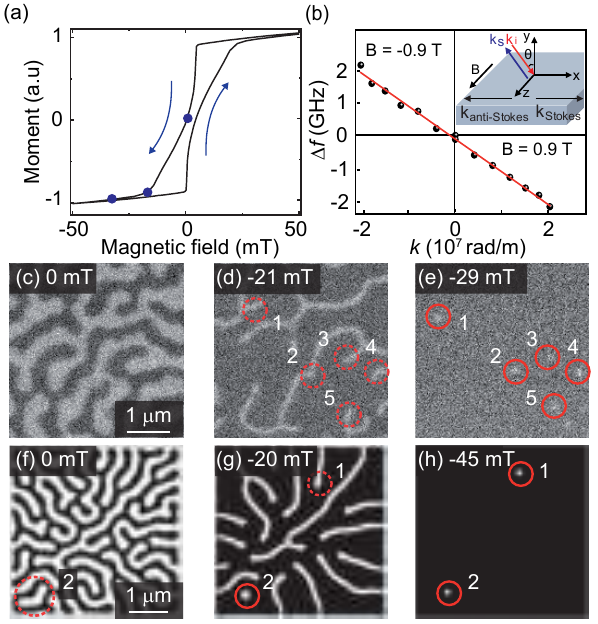} 
\caption{
(a) Out of plane hysteresis loop of the [Pt/Co/Ta]$_{12}$ film. (b) The frequency asymmetry ($\Delta f$) extracted from the BLS measurement of the [Pt/Co/Ta]$_{12}$ film for an in-plane applied bias field of 0.9 T and -0.9 T. $\Delta f$ is the difference between the frequency of the Stokes and the anti-Stokes peaks present in the BLS spectra. A schematic of the experimental configuration is shown in the inset. 
(c-e) XMCD-STXM of magnetic states in the multilayer film at thee different magnetic fields indicated with blue points on the hysteresis loop in (a). (f-h) Simulated images of similar magnetic configurations in the multilayer film. In all images, the dark and bright contrast corresponds to domains with magnetization pointing up and down, respectively. The skyrmions are indicated by solid red circles and the domain ends from which the skyrmions are nucleated are indicated by dashed red circles.
}
\label{fig2}
\end{figure}

BLS measurements were carried out by focusing a monochromatic laser ($\mathrm{\lambda}$ = 532 nm) on the sample surface and analyzing the backscattered light by a Sandercock-type (3+3) pass tandem Fabry-Perot interferometer. An in-plane magnetic field, sufficiently large to saturate the magnetization in the film plane, was applied along the direction perpendicular to the wavevector of the spin waves. Due to the photon-magnon conservation law of momentum in the scattering process, the magnitude of the wave vector is linked to the incidence angle of light $\theta$, by the relation $k = 4 \pi \sin{\theta / \lambda}$. For our measurements, $\Delta f$ measured as a function of $k$, for an applied bias field ($B$) of 0.9 T and -0.9 T. $k$ was varied from 0 to $2\times10^{7}$ rad/m and a linear behavior is observed. Note that, upon reversing the direction of the applied magnetic field, $\Delta f$ changes sign due to the reversal of the spin wave propagation direction. The effective DMI constant was then determined with a linear fit (continuous line) to the experimental data based on Eq. 1.
The value of DMI, $D$ = - 0.78 $\pm ~0.02$ mJ/$\mathrm{m}^{2}$ was determined using gyromagnetic ratio $\Gamma$ = 19.4 GHz/T and previously measured value of $M_{\mathrm{s}}$. the value of DMI is sufficient to support the formation of N\'eel skyrmions. Moreover, the negative value of $D$ indicates that left-handed chirality is favored in this material system.
\newline \indent The magnetic configuration of the antidot lattices was imaged using STXM at the PolLux endstation of the Swiss light source~\cite{Raabe2008}. Circularly polarized photons are tuned to the L$_{3}$ absorption edge of Co (778 eV) and are focused onto a 25 nm spot on the sample using a diffractive Fresnel zone plate with an outermost zone width of $\approx$ 25 nm. Magnetic contrast was obtained through the x-ray magnetic circular dichroism (XMCD) effect and, by illuminating the sample at normal incidence with respect to the surface. A selection of XMCD images acquired under different applied out-of-plane magnetic fields of the magnetic states in a continuous [Pt/Co/Ta]$_{12}$ multilayer film are shown in Fig. 2 (c-e). Prior to the acquisition of the images, the following field initialization protocol was applied; the sample was saturated with a 100 mT out-of-plane magnetic field, which was then reduced to 6 mT. After the field initialization, a demagnetized sample with labyrinth domains was observed. These demagnetized labyrinth magnetic states are always obtained after repeated field initialization protocols. Then the magnetic field reduced to zero and a reversed field was applied in increasing steps. As the magnetic field is increased up to the reverse saturation field, the width of the labyrinth domains with magnetization pointing antiparallel to the applied field is reduced from a value of $\approx 200 \pm ~35$ nm at -6 mT to $\approx 150 \pm ~35$ nm upon their collapse into several isolated skyrmions at -23 mT. Close inspection reveals that the skyrmions form at the end of the domains. These skyrmions eventually annihilate at an applied magnetic field of $\approx-38$ mT.

\section{Micromagnetic simulations of Skyrmion nucleation}

 The skyrmion nucleation process in a thin film and antidot lattices are simulated using Mumax3~\cite{MuMax2011_main,mumax_2014,Leliaert2014}. For the simulation, we have used periodic boundary conditions along the out-of-plane axis with six repetitions in both the negative and positive $z$ directions~\cite{woo2016observation}. The simulations were performed for a finite $\SI{4.4}{\micro\metre}^2$ film using a uniformly discretized grid with a cell size of 2.0 nm $\times$ 2.0 nm $\times$ 1.1 nm, where the thickness of the film cell of 1.1 nm corresponds to the thickness of the magnetic layer in the real sample. For the simulations, we employed the experimentally measured values of $M_\mathrm{s}$ and DMI reported in Section III. Furthermore, an exchange constant of $A = 1 \times 10^{-11}$ J/m and a PMA of $K_\mathrm{u}=3.45 \times 10^{5}$~J/$\mathrm{m^{3}}$, taken from the literature~\cite{Stoop2018CloggingLandscapes}, were used in the simulations. In order to mimic the polycrystalline nature of the sample, a Voronoi tessellation was added in the micromagnetic simulations, where each grain was assigned a slightly different perpendicular magnetic anisotropy drawn randomly from a normal distribution centred about a mean value with a standard deviation of $10\%$. The  variation of anisotropy is important to reproduce the experimental observations in simulation as it acts as defects, which facilitating the skyrmion formation~\cite{Yu2012MagneticReversals}. 
For continuous thin films, the ratio of magnetization pointing down domain (black) to magnetization pointing up (white) domains is controlled by the Zeeman energy, the size and morphology of the black and white domains are controlled by the anisotropy and domain wall energy, and the chirality is determined by the DMI. The experimental observation and simulations of the continuous thin film [Fig. 2 (f-g)] exhibit a good qualitative agreement with the experimental images shown in Fig. 2 (c-e). In the simulations, a labyrinth domain pattern is formed at zero magnetic field [Fig. 2 (f)]. As the magnetic field was increased, the labyrinth domains collapsed into individual skyrmions [Fig. 2 (d,e and g, h )]. The skyrmions are formed at the end of the white domains, which is similar to the experimental observations. However, in the simulations, the magnetic skyrmions were created at a larger magnetic field than that observed in the experimental data. This observed discrepancy between the micromagnetic simulations and the experimental data can be explained by the fact that the simulations were performed at 0 K, neglecting any thermal fluctuations, which leads to an increase of the nucleation field ~\cite{Suess2011CalculationTemperatures,Sharrock1994Timeinvited}. 

In order to provide insight into the skyrmion stabilization and capture in the antidot lattices, micromagnetic simulations were systematically performed for lattice constants of 400, 500, 600 and 800 nm. The evolution of the domain patterns on increasing the applied magnetic field is shown in Fig. 3. In general, during the magnetization reversal of the antidot lattice, reverse domains (in black) penetrate into the antidot lattice through the channels between the holes. The penetration of the black domains is hindered by the pinning field associated with the antidots, and white domains are pinned around the holes. 

\begin{figure*}[!htp]
\includegraphics[scale=1,width=0.9\textwidth]{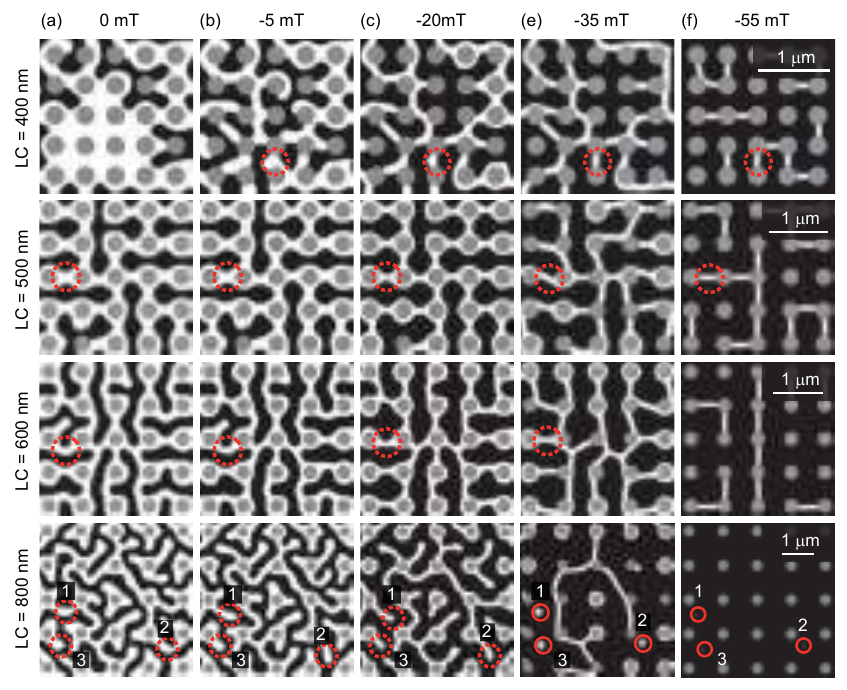} 
\caption{
Simulated images of the evolution of the magnetic configuration in antidot lattices with different lattice constants on increasing the magnetic field. The dark and bright contrast corresponds to domains with magnetization pointing up and down, respectively. Pinned domains are indicated by red dashed circles and the skyrmions are marked with circles with red solid lines.
}
\label{fig3}
\end{figure*}

\indent For the more compact lattices (LC = 400, 500 and 600 nm), the labyrinth domain pattern is significantly modified at zero field [see, for example, Fig. 3(a)] and the details of the magnetization reversal strongly depends on the lattice constant. At 0 mT, while the black domains have penetrated into the antidot lattice for the larger lattice constants, for LC = 400 nm, most of the lattice is still in a saturated state. A higher magnetic field is then required for the black domains to penetrate into the lattice and, due to the lack of space between the antidots, the white domains do not completely surround the holes. For all three lattice constants, the white domains often remain pinned at two neighboring holes on increasing the magnetic field, forming straight narrow domains between them, an example of which is indicated by red dotted circles in Fig. 3 for LC = 400, 500 and 600 nm. With further increase of the magnetic field, these connected white domains shrink to form straight sections between the antidots and eventually vanish at a higher magnetic field. As a result, for these compact lattices, the formation of the skyrmions is hindered due to strong pinning of the labyrinth domains at the holes in the antidot lattice. 

\indent For sufficiently large lattice constant, LC = 800 nm, a partial labyrinth domain pattern is observed at 0 mT. In this case, some of the domains are either directly connected between two neighboring holes [see, for example, dashed circles labelled 1 and 2 in Fig. 3 (a)] but also form open domain ends [see dashed circle labelled 3 in Fig. 3 (a)]. As the applied magnetic field is increased, the connecting domains disconnect, forming open domain ends as indicated by the dashed circles labelled 1 and 2 in Fig. 3. With further increase of the applied magnetic field, these open domain ends in the antidot lattice eventually shrink to form magnetic skyrmions indicated by red solid circles in Fig. 3 (e) and (f). The size of the skyrmions also decreases with the further increase of the magnetic field [compare, for example, skyrmions in Fig. 3 (e) and (f), and see Supplementary movie 2 for LC = 800 nm].  The detailed evolution of the 3D magnetization configuration on increasing of the applied magnetic field for LC = 400 and 800 nm is shown in Supplementary movies 3 and 4, respectively.

\begin{figure*}[!htp]
\includegraphics[scale=1,width=0.85\textwidth]{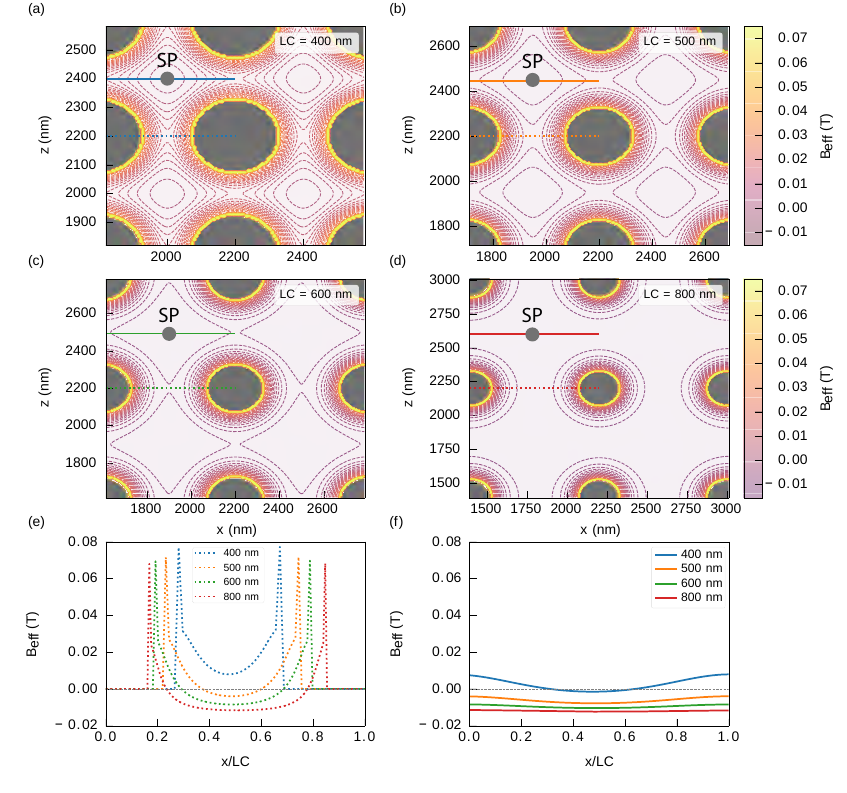}  
\caption{
(a-d) Contour plots of the simulated effective field distributions with the colour map are shown for different lattices at 5 mT. (e-f) Simulated effective field from different lattices obtained by taking a line scan in two different regions marked by horizontal dashed and solid line in Fig. 4 (a-d), respectively. Individual colors represent individual lattices.
}
\label{fig4}
\end{figure*}

In order to understand the magnetization reversal process and the mechanism behind the skyrmion formation in the antidot lattice, we have extracted the spatial distribution of the effective magnetic field from the simulations for the different lattice constants at a field value of 5 mT so that the system is in a saturated state just before the domains formation. From Fig. 4 (a-d), it can be seen that the contour lines of the effective field are very dense around the holes and the lines density between the holes decreases with increasing the lattice constant. The profile of effective field along the dashed and solid lines in Fig. 4 (a-d) are shown in Fig. 4 (e) and (f), respectively. Form this, it becomes clear that the effective field profile changes, in general decreasing and flattening, with the increase of the lattice constant in both the regions. The magnetization reversal process is facilitated or hindered, depending on whether the out-of-plane component of the effective field is negative or positive, respectively. 
Although the change in the effective field along the profiles is relatively small, it is sufficient to modify the domain patterns. In particular, in Fig. 4 (e), it can be seen that the effective field has a high positive value close to the edges of the holes ~\cite{Kos2012, Kruglyak_2017} which explains the pinning of the white domains around the holes. Additionally, the effective field is a minimum at the saddle point (SP), which explains the formation of black domains at the saddle point region, particularly noticeable for LC = 500 and 600 nm. The larger effective field in the region between neighboring holes supports the pinning white domains forming a connection between the holes. For an antidot lattice with LC = 800 nm, the effective field along the channels between the holes (along the solid line in Fig. 4) is almost uniform and, in order to find an equilibrium state, the system forms labyringth domain patterns with open domain ends. As the external magnetic field changes, the domain width also changes to minimize the total energy of the system and, as shown in literature \cite{Wolfe2013AppliedResearch}, the length of the domain will also decrease to minimize the total energy. During the contraction of the domains, the open domain end is pinned at a local region with high anisotropy and, subsequently, a constriction forms that narrows until a skyrmion is formed. This mechanism of formation of skyrmion resembles the necking and detachment of a drop at the edge of a circular capillary tube ~\cite{Zhang1997DropTube}. We repeated the simulations several times for different variations in the anisotropy field distribution. The simulated results showed that, while the formation of skyrmion process is always the same, both the number of skyrmions and their position strongly depend on the anisotropy distribution.

\begin{figure*}[!htp]
\includegraphics[scale=1,width=0.9\textwidth]{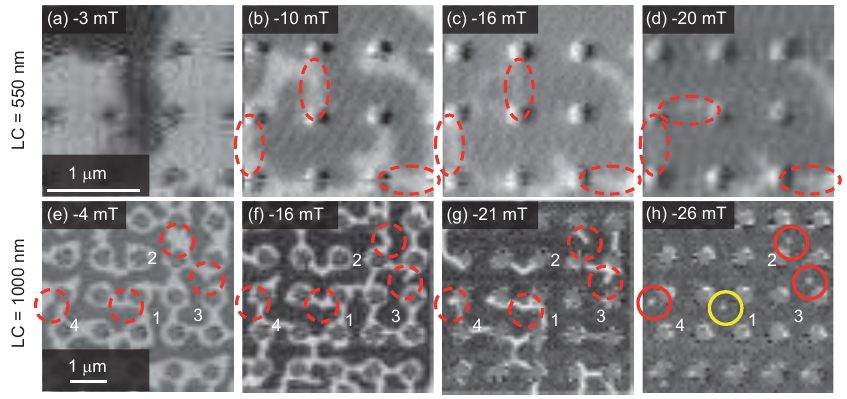}
\caption{
XMCD-STXM images of the evolution of the magnetic configuration in the antidot lattices with magnetic field for anitdot lattices with (a-d) LC = 550 nm and (e-h) LC = 1000 nm. The dark and bright contrast corresponds to magnetic domains with magnetization pointing up and down, respectively. The skyrmions are indicated by solid circles and the domains from which the skyrmions are nucleated are indicated by red dashed circles and ellipses. 
}
\label{fig5}
\end{figure*}

\section{Experimental observations}

Following the micromagnetic simulations, we performed experiments on two samples patterned into antidot lattices with different lattice constants, LC = 550 nm and 1000 nm. The XMCD-STXM images of the domain structure in the antidot lattices are shown in Fig. 5 (a-h). As for the experiments on the continuous [Pt/Co/Ta]12 films, we first performed a field initialization protocol, decreasing the field from 100 mT to zero and then increasing the magnetic field in reverse direction. For the antidot lattice with LC = 550 nm, the labyrinth domains form connections between the antidots on increasing the magnetic field as indicated by red dashed ellipses in Fig. 5 (b, c and d), which prevents the formation of skyrmions. In contrast, for LC = 1000 nm, initially the labyrinth domains directly connect two neighboring antidots as shown, for example, by red dashed circles labelled 1 and 2 in Fig. 5 (e, f and g). With increase in magnetic field, the domains disconnect from one of the antidots creating a "tail" as shown by the red dashed circles labelled 2, 3 and 4 in Fig. 5 (g). The domains with open ends progressively shrink down in width and length until they collapse into isolated magnetic skyrmions indicated by red and yellow solid circles in Fig. 5 (h). This experimental observation that skyrmions only form for larger lattice constants is consistent with the results of the micromagnetic simulations, in which it was observed that skyrmions only form for a lattice constant of 800 nm or larger. In particular, we note that the mechanism for the skyrmion formation is well-reproduced by the simulations. 
\begin{figure}[ht!]
\includegraphics[scale=1,width =\columnwidth]{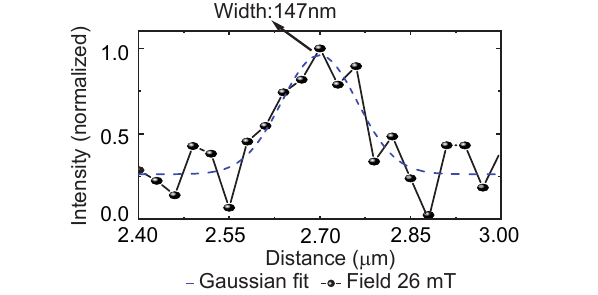} 
\caption{
Line profile of the magnetic contrast across the skyrmion marked with yellow solid circle In Fig. 4 (h). The profiled is fitted with a Gaussian curve and has a full width half maximum of 147 nm.
}
\label{fig5}
\end{figure}

The domains marked by red dashed circles in Fig. 5 collapse into isolated skyrmions at an applied field of about -22 mT. The diameter of one of the skyrmions (indicated by the yellow circle in Fig. 5(e)] was determined to be 147 $\pm~35$ nm (see Fig. 6). This is comparable with the simulated skyrmion size, which decreases from 200 nm to 100 nm as the field is increased from -35 mT to -55 mT although, in the experiment, no significant change in the skyrmion size on increasing the magnetic field could be seen. The skyrmions finally annihilate at a higher applied field and, on repeating the initialization protocol described earlier, we observed that the magnetic skyrmions can be repeatedly created in the antidot lattice. The experiment and simulation both show that unconnected domain ends are a necessary condition for the formation of the magnetic skyrmions in the antidot lattice. 

\section{Summary}
We have elucidated the mechanism for the formation of N\'eel skyrmions in an antidot lattice combining micromagnetic simulations and XMCD-STXM measurements. Micromagnetic simulations have revealed two distinct regimes, one characterized by small lattice constant $\mathrm{LC} \leq$ 600 nm that suppresses the formation of skyrmions, while skyrmions are formed for antidot lattice with $\mathrm{LC} \geq$ 800 nm. Therefore, the lattice constant of the antidot array plays a pivotal role in the skyrmion formation. In particular, the introduction of the holes in a thin magnetic film modifies the spatial distribution of the effective field, which has a profound influence on the formation mechanism of the skyrmions. For a defect-free antidot lattice, the skyrmions will formed at the saddle point as a result of the spatial distribution of the effective field. In our system, which has a variation in the local anisotropy, the position of skyrmion formation is determined by the region with the highest anisotropy value.  The mechanism of formation of skyrmions inside the antidot lattice resembles the formation of liquid droplet at the edge of a capillary tube. On the increasing in applied magnetic field, a constriction forms and subsequently narrows until a skyrmion is formed, which is equivalent to necking and detachment of the liquid drop.
In order to benchmark the micromagnetic simulation with a real system, we used focused ion beam milling to fabricate magnetic antidot lattices with LC = 550 and 1000 nm in a multilayered [Pt/Co/Ta$]_{12}$ film with perpendicular magnetic anisotropy and imaged the evolution of the magnetic configuration in presence of an applied magnetic field with scanning transmission x-ray microscopy. Our experimental results reflect the prediction obtained by the micromagnetic simulation. Not only do we observe the formation of N\'eel skyrmions for the sample with larger lattice constant (LC = 1000 nm), but also the size of the skyrmions is comparable with the predictions of the micromagnetic simulation. 
In conclusion, the array of artificial pinning sites introduced in the form of an antidot lattice can host skyrmions, which form in arrays with sufficiently large lattice constant. We envisage that the effective field resulting from these pinning sites can be used to control their position, making the antidot lattice a promising candidate for control of current driven skyrmion motion.

\section{Acknowledgments}
We would like to acknowledge Dr. Peter Derlet, Staff Scientist at PSI for his valuable scientific inputs. XMCD-STXM measurements were performed at the PolLux endstation installed at the X07DA beamline of the Swiss Light Source, Paul Scherrer Institute, Switzerland. We thank Dr. E. Müller and A. Weber for sample fabrication. S.S. acknowledges ETH Zurich Post-Doctoral fellowship and Marie Curie actions for People COFUND program (Grant No. FEL-11 16-1). M.M. acknowledges funding from the EU FP7 SASPRO Programme (REA Grant Agreement No. 609427, project WEST 1244/02/01) with further co-funded by the Slovak Academy of Sciences. A.K.S. acknowledges funding from the European Community's Seventh Framework Program (FP7/ 2007-2013) under Grant No. 290605 (COFUND: PSI-FELLOW). M.Z. and M.K. acknowledge funding from the EU Horizon 2020 project MagIC Grant No. 644348, Polish Ministry of Science and Higher Education resources for science in 2017-2019 (W28/H2020/2017), National Science Centre of Poland Grant No. 2017/27/N/ST3/00419 and Adam Mickiewicz University Foundation. The simulations were partially performed at the Poznan Supercomputing and Networking Center (Grant No.~398). Financial support by the EMPIR Programme under the Grant Agreement 17FUN08 TOPS. S.F. acknowledges funding from the EU Horizon 2020 project MagicSky (Grant No. 665095). The PolLux end station was financed by the German Ministerium für Bildung und Forschung (BMBF) through contracts 05KS4WE1/6 and 05KS7WE1. S.T. acknowledges financial support by the European Metrology Programme for Innovation and Research (EMPIR), under the Grant Agreement 17FUN08 TOPS.

\bibliographystyle{apsrev4-1}

\end{document}